\documentclass[12pt,preprint]{aastex}
\usepackage{graphicx, amsmath, amssymb}
\usepackage[dvips]{color}
\usepackage{lineno} % to include line numbering functionality for collaborators

\newcommand{\pow}[2]{\ensuremath{\mbox{#1}^{\rm #2}}}
\newcommand{\subs}[2]{\ensuremath{\mbox{#1}_{\rm #2}}}
\newcommand{\del}[1]{\ensuremath{\Delta #1}}
\newcommand{\Blos}{\ensuremath{B_{\rm los}}}

\newcommand{\methanol}{CH$_3$OH}
\newcommand{\kmS}{km \pow{s}{-1}}
\newcommand{\cm}[1]{\pow{cm}{#1}}    %% use \cm{power}
\newcommand{\mG}{\pow{mG}{-1}}

\shortauthors{Sarma \& Momjian}
\shorttitle{EVLA Zeeman Effect Discovery in the 44 GHz CH$_3$OH maser}

\begin{document}

\title{Discovery of the Zeeman Effect in the \\ 
	44 GHz Class I methanol (CH$_3$OH) maser line}

\author{A.~P.\ Sarma\altaffilmark{1}, E.\ Momjian\altaffilmark{2}}

\altaffiltext{1}{Physics Department, DePaul University, 
2219 N. Kenmore Ave., Byrne Hall 211, 
Chicago IL 60614; asarma@depaul.edu}

\altaffiltext{2}{National Radio Astronomy Observatory, Socorro NM 87801}

\begin{abstract}
We report the discovery of the Zeeman effect in the 44 GHz Class I methanol (\methanol) maser line. The observations were carried out with 22 antennas of the EVLA toward a star forming region in OMC-2. Based on our adopted Zeeman splitting factor of $z = 1.0$ Hz \mG, we detect a line of sight magnetic field of $18.4 \pm 1.1$ mG toward this source. Since such 44 GHz \methanol\ masers arise from shocks in the outflows of star forming regions, we can relate our measurement of the post-shock magnetic field to field strengths indicated by species tracing pre-shock regions, and thus characterize the large scale magnetic field. Moreover, since Class I masers trace regions more remote from the star forming core than Class II masers, and possibly earlier phases, magnetic fields detected in 6.7 GHz Class II and 36 GHz and 44 GHz Class I methanol maser lines together offer the potential of providing a more complete picture of the magnetic field. This motivates further observations at high angular resolution to find the positional relationships between Class I and Class II masers, and masers at various frequencies within each category. In particular, \methanol\ masers are widespread in high- as well as intermediate-mass star forming regions, and our discovery provides a new method of studying the magnetic field in such regions, by observing small physical scales that are not accessible by any other lines.
\end{abstract}


\section{INTRODUCTION}
\label{sINTRO}

%\linenumbers % to turn on line numbering

The presence of magnetic fields in star forming regions presents theoretical and observational challenges because of the computational complexity involved in incorporating magnetic fields in simulations, and  the difficult observations that are required to measure their strengths and directions (\citealt{tt2008}). While the role of magnetic fields in regulating the onset of star formation is still a matter of debate (e.g., \citealt{rmc2009}), it has become increasingly clear that magnetic fields play a critical role in carrying angular momentum away from the protostar during collapse (\citealt{mckee2007}, and references therein). Moreover, the outflows along which this takes place may be driven by dynamically enhanced magnetic fields in the protostellar disk (\citealt{bp2006}). In particular, the driving of outflows along magnetic field lines may be critical in allowing accretion to continue onto high mass protostars (\citealt{bp2007}). Observational studies of such protostellar disks and outflows are difficult, due to the need for high angular resolution. Interstellar masers, being compact and intense, offer a means of probing regions at high angular resolution. For years, their effectiveness as probes of star forming regions was overshadowed by a perception that the specialized conditions in which such masers form necessarily prevented them from being linked to conditions on larger scales. However, recent discoveries indicate that rather than being a measure in isolated atypical fragments, the magnetic fields measured in masers are indeed linked to the larger scale magnetic field. \citet{fr2006}\ found a relative consistency in the magnetic fields measured in clusters of mainline (1665 and 1667 MHz) OH masers across a massive star forming region, and concluded that magnetic fields are ordered in massive star forming regions. More recently, \citet{wv2010}\ have determined from polarization observations of 6.7 GHz \methanol\ masers toward Cepheus HW2 that the masers probe the large scale magnetic field. 

The 6.7 GHz \methanol\ masers mentioned above are examples of Class II methanol masers, which are known to be probes of the early phases of high mass star forming regions (\citealt{wv2008}, and references therein). Yet another class of masers, namely Class I methanol masers, may probe even earlier phases of star forming regions (\citealt{pp2008}). Such Class I \methanol\ masers are believed to be formed in collisional shocked regions in protostellar outflows (\citealt{cragg1992}; \citealt{san2005}). Together, Class I and Class II \methanol\ masers offer a unique window into star forming regions by allowing us to observe 
the smallest physical scales that are not accessible via any other lines. Given the value of measuring the magnetic field in disks and outflows, we have embarked on a long-term plan to measure magnetic fields in \methanol\ masers. In 2009, we discovered the Zeeman effect in the 36 GHz Class I \methanol\ maser line toward the high mass star forming region M8E (\citealt{sm2009}). The Zeeman effect remains the most direct method for measuring magnetic field strengths (e.g., \citealt{rmc99}). 

In this paper, we report the \textsl{discovery} of the Zeeman effect in the 44 GHz Class I \methanol\ maser line. The observations were carried out with 22 antennas of the Expanded Very Large Array (EVLA) toward OMC-2. OMC-2 is a filamentary structure containing several sites of active star formation (e.g., \citealt{taka2008}), and located about 12$\arcmin$ northeast of the Trapezium OB cluster in Orion (\citealt{cas1995}), at a distance of 450 pc (\citealt{gs1989}). In \S\ \ref{sODR}, we present details of the observations and reduction of the data. The analysis involved in extracting magnetic field information from the Zeeman effect is given in \S\ \ref{sANAL}. The results are presented and discussed in \S\ \ref{sR}.   

\section{OBSERVATIONS AND DATA REDUCTION}
\label{sODR}

Observations of the $7_{0}-6_1\, A^+$ methanol maser emission line at 44 GHz were carried out using 22 antennas in the D-configuration of the Expanded Very Large Array (EVLA) of the NRAO\footnote{The National Radio Astronomy Observatory (NRAO) is a facility of the National Science Foundation operated under cooperative agreement by Associated Universities, Inc.} in two 2 hr sessions on 2009 Oct 25 and Nov 25. Table\ \ref{tOP}\ lists the observing parameters and other relevant data for these observations. The data were correlated using the old VLA correlator, and in order to avoid the aliasing known to affect the lower 0.5 MHz of the bandwidth for EVLA data correlated with the old VLA correlator, the spectral line was centered in the second half of the 1.56 MHz wide band. The source 3C147 (J0542$+$4951) was used to set the absolute flux density scale, while the compact source J0607$-$0834 was used as an amplitude calibrator.

The editing, calibration, Fourier transformation, deconvolution, and 
processing of the data were carried out using the Astronomical Image Processing 
System (AIPS) of the NRAO. After applying the amplitude gain corrections of 
J0607$-$0834 on the target source OMC-2, the spectral channel with the strongest 
maser emission signal was split, then self-calibrated in both phase and amplitude 
in a succession of iterative cycles (e.g., \citealt{st2002}). The final phase and amplitude solutions were 
then applied to the full spectral-line $uv$ data set, and Stokes $I$ and $V$ image 
cubes were made with a synthesized beamwidth of $1.93\arcsec \times 1.58\arcsec$; the maser emission was unresolved in these observations. 
Further processing of the data, including magnetic field estimates, 
was done using the MIRIAD software package.

\section{ANALYSIS}
\label{sANAL}

For cases in which the Zeeman splitting \subs{\del{\nu}}{z}\
is much less than the line width \del{\nu}, the magnetic field
can be obtained from the Stokes $V$ spectrum, which exhibits a
{\em scaled derivative} of the Stokes $I$ spectrum (\citealt{hgmz93}). 
Here, consistent with AIPS conventions,
$I$ = (RCP$+$LCP)/2, and $V$ = (RCP$-$LCP)/2; RCP is right- and 
LCP is left-circular polarization incident on the antennas, where
RCP has the standard radio definition of clockwise rotation of the 
electric vector when viewed along the direction of wave propagation. 
Since the observed $V$ spectrum may also contain a scaled replica 
of the $I$ spectrum itself, the Zeeman effect can be measured by 
fitting the Stokes $V$ spectra in the least-squares sense to the  equation   
\begin{equation}
\mbox{V}\ = \mbox{aI}\ + \frac{\mbox{b}}{2}\ 
\frac{\mbox{dI}}{\mbox{d}\nu}
\label{eVLCO}
\end{equation}
(\citealt{th82}; \citealt{skzl90}). The fit parameter $a$ is usually the result of small calibration errors in RCP versus LCP, and is expected to be small. In these observations, $a$ was of the order of $10^{-4}$ or less. While eq.~(\ref{eVLCO})\ is strictly true only for thermal lines, numerical solutions of the equations of radiative transfer (e.g., \citealt{nw92}) have shown that it gives reasonable values for the magnetic fields in masers also. In eq.~(\ref{eVLCO}), the fit parameter $b = zB$\,cos\,$\theta$, where $z$ is the Zeeman splitting factor (Hz \mG), $B$ is the magnetic field, and $\theta$ is the angle of the magnetic field to the line of sight (\citealt{ctg93}). The value of the Zeeman splitting factor $z$ for \methanol\ masers is very small, because \methanol\ is a non-paramagnetic molecule. Following the treatment of \citet{wv2008} for the Zeeman splitting of 6.7 GHz methanol masers, we derive the Zeeman splitting factor using the Land$\acute{e}$ $g$-factor based on laboratory measurements of 25 GHz methanol masers (\citealt{jen51}), and find $z$ = 1.0 Hz \mG. Since the actual Land$\acute{e}$ $g$-factor for the 44 GHz line is not known, it is not possible to estimate the uncertainty in $z$. It is unlikely, however, that the measured Land$\acute{e}$ $g$-factor will be vastly different from the value used here, since the \Blos\ values derived below seem to agree well with projections about the field and densities (see \S\ \ref{sR} below). In fact, if the measured value of the Land$\acute{e}$ $g$-factor turns out to be vastly different, the Zeeman interpretation of these observations may have to be revised. Needless to say, these observations and the above considerations motivate immediate measurement of the Land$\acute{e}$ $g$-factor in the 44 GHz \methanol\ maser line.

\section{RESULTS AND DISCUSSION}
\label{sR}

\begin{figure}
\centering
\includegraphics[width=0.9\textwidth]{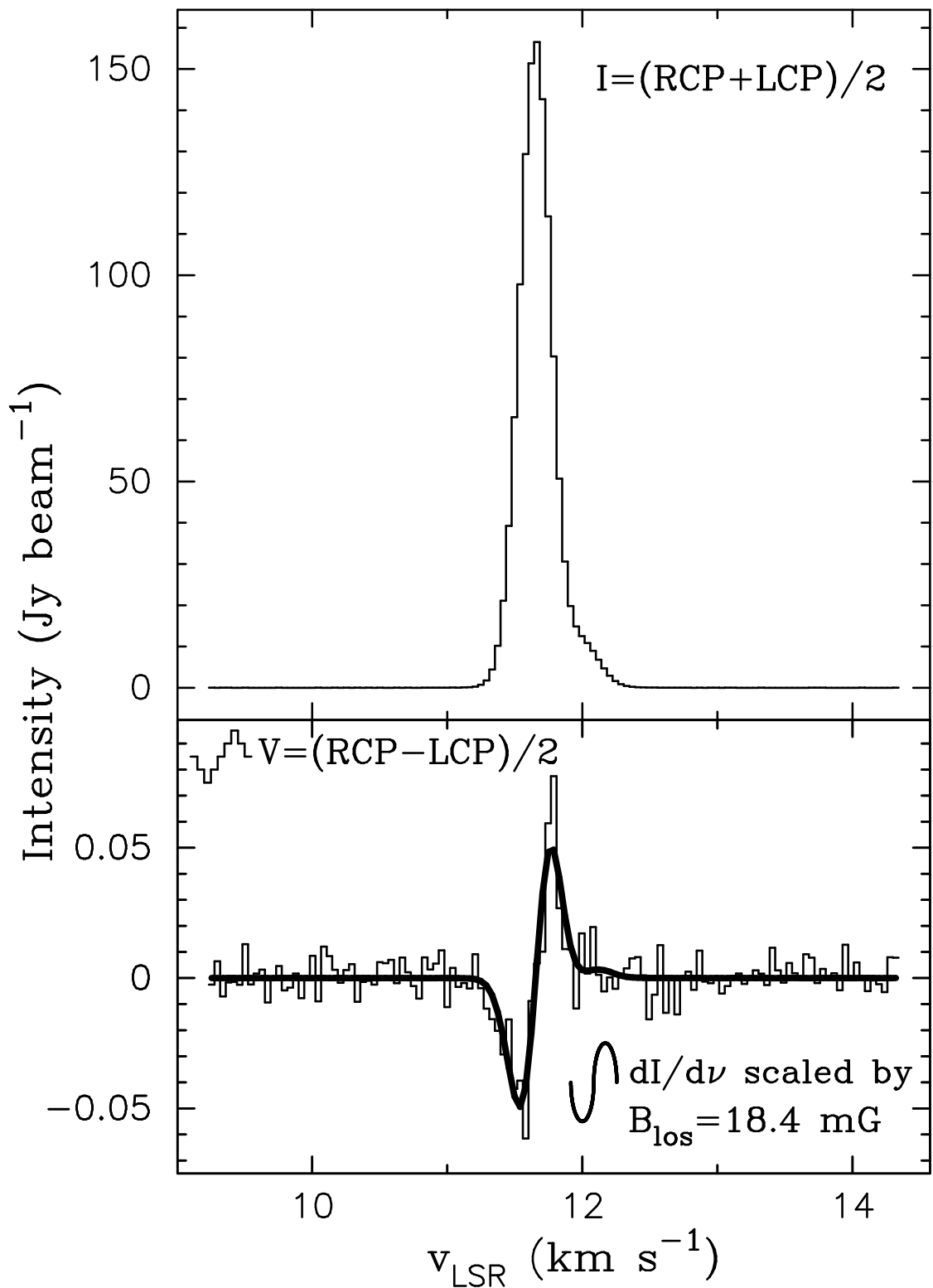}
%\plotone{./FIGS/omc2_ivd.ps}
%\plottwo{./FIGS/f1a.eps}{./FIGS/f1b.eps}
\caption{Stokes $I$ ({\em top$-$histogram}) and $V$ ({\em bottom$-$histogram})
profiles of the maser toward the 44 GHz Class I \methanol\ maser in OMC-2.
The curve superposed on $V$ in the lower frame is the derivative of $I$ scaled 
by a value of \Blos\ = $18.4 \pm 1.1$ mG. \label{fIVD}}
\end{figure}

Figure \ref{fIVD}\ shows our Zeeman effect detection in the 44 GHz \methanol\ maser toward OMC-2. As described in \S\ \ref{sANAL}, we determined magnetic fields by fitting the Stokes $V$ spectra in the least-squares sense using equation (\ref{eVLCO}). The fit parameter $b$ (see eq.~\ref{eVLCO}) obtained by this procedure, together with the Zeeman splitting factor $z = 1.0$ Hz \mG\ discussed in \S\ \ref{sANAL}\ above, then gives \Blos\ = $18.4 \pm 1.1$ mG for the 44 GHz Class I \methanol\ maser toward OMC-2. By convention, a positive value for \Blos\ indicates a field pointing away from the observer.

\begin{figure}
\centering
\includegraphics[width=0.8\textwidth]{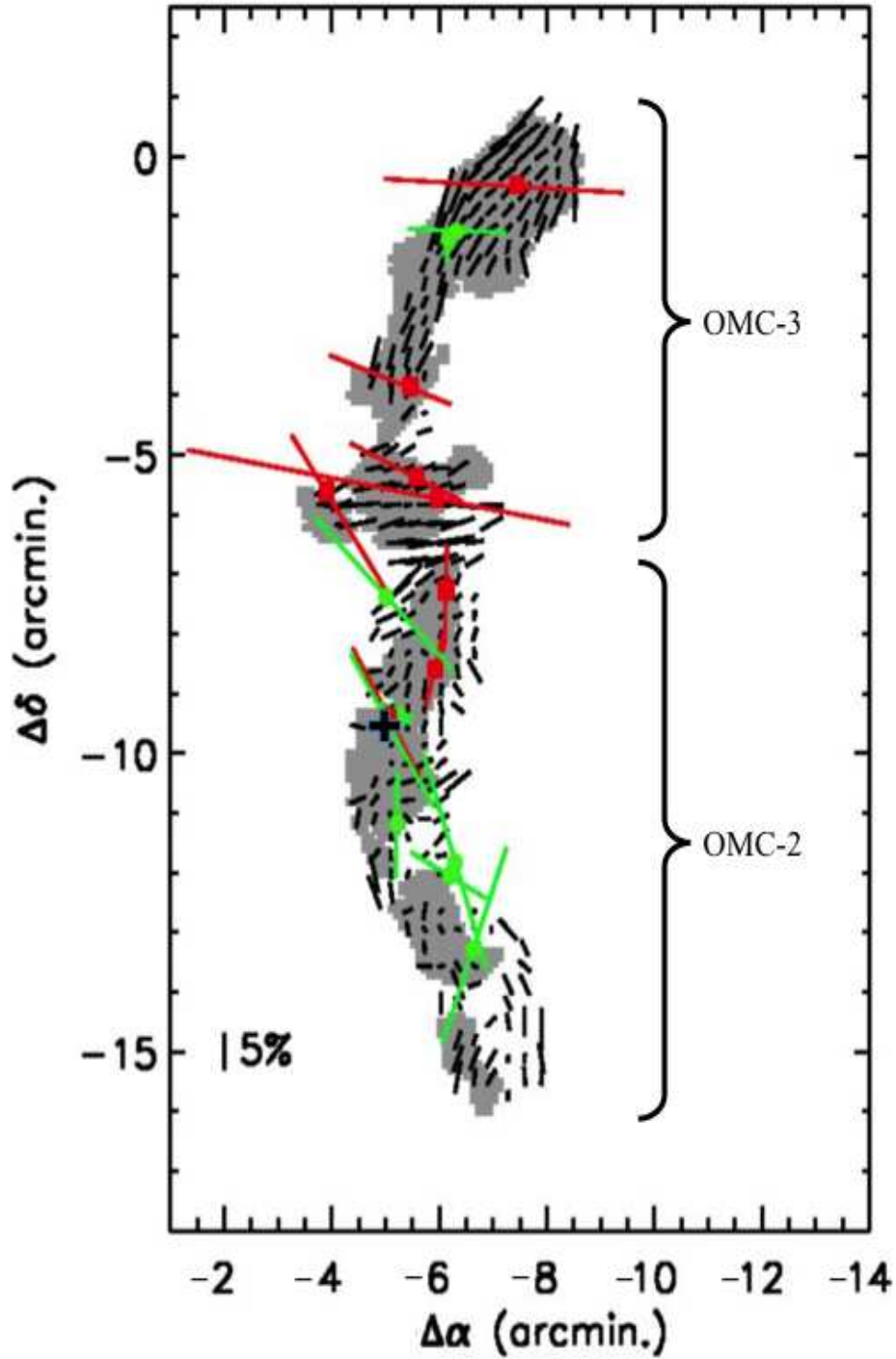}
%\plotone{./FIGS/omc2_ivd.ps}
%\plottwo{./FIGS/f1a.eps}{./FIGS/f1b.eps}
\caption{Distribution of H$_2$ jets (green) and CO outflows (red) in OMC-2 and OMC-3, superimposed on 850 $\mu$m polarization data (thin vectors), taken from Figure 7 of \citet{pbm2010}. The greyscale is the 850 $\mu$m intensity, and is displayed to show the location of the filament. The reference position is $\alpha_{2000} = 05^\text{h} 35^\text{m} 48^\text{s}, \delta_{2000} = -05\arcdeg00\arcmin0\arcsec$. The thick black cross shows the location of our 44 GHz \methanol\ maser. \label{f850}}
\end{figure}

In order to investigate the connection of our detected field to quantities in the larger region, we looked at the 850 $\mu$m polarization map of \citet{pbm2010} observed with the JCMT (James Clerk Maxwell Telescope). Their map is reproduced in Figure \ref{f850}. For reference, the position of our \methanol\ maser is marked in this figure. We have used the position of the strongest 44 GHz \methanol\ maser from the high angular resolution phase referenced VLA A-configuration observations (HPBW 0$\arcsec$.15) of \citet{sk2009}. Associated with this position is a CO outflow (\citealt{taka2008}; \citealt{wph03}) marked in red in Figure \ref{f850}; the position angle of this CO outflow is 30$^\circ$. Also shown in green in Figure \ref{f850}\ is the outflow detected in the 2.12 $\mu$m %$v = 1-0$  S(1) 
shocked H$_2$ emission line from \citet{stanke2002}; the position angle of this outflow is $\sim$31$^\circ$, identical to that of the CO outflow. In their high angular resolution VLA observations referred to above, \citet{sk2009}\ observed six 44 GHz methanol masers spread out along a line aligned at an angle approximately 30$^\circ$ east of north. Upon comparing all of the above information to the 850 $\mu$m polarization map of \citet{pbm2010}, it appears likely that the outflow is aligned along the local magnetic field, and the methanol masers are excited in collisional shocks along the outflow. Maser amplification is particularly efficient in directions approximately perpendicular to the shock propagation, and the compression of an ordered magnetic field from this orientation would give the following relationship:
\begin{equation}
\frac{B_0}{\rho_0} = \frac{B_1}{\rho_1}   \label{e1}
\end{equation}
where 0 and 1 refer to the preshock and postshock (maser) regions respectively, and $\rho$ is the gas density (e.g., \citealt{sarma2008}).  In the discussion below, we will use the molecular hydrogen number density $n$ instead of $\rho$. If $B_0$ and $n_0$ are known, then we can use our detected value of \Blos\ to find $n_1$, and compare it to the density at which the 44 GHz \methanol\ maser is excited. \citet{pbm2010}\ used the Chandrasekhar-Fermi method to estimate the value of the magnetic field in this region from their 850 $\mu$m observations, and found it to be $0.13$ mG; we adopt this as our value of $B_0$. We will use $n_0 = 10^4$ \cm{-3}, which was obtained by \citet{cas1995} based on C$^{18}$O observations toward OMC-2 (also see  \citealt{pbm2010}). Finally, we adopt $B_1 = 2 \Blos$, based on statistical grounds (\citealt{rmc99}). Using these values in equation (\ref{e1}), we get $n_1 \sim 10^6$ \cm{-3}\ for the postshock number density. This is in excellent agreement with theoretical models that show the 44 GHz \methanol\ maser action is maximized in regions with density $10^5-10^6$ \cm{-3}\ (\citealt{pp2008}, and references therein). Another way of looking at this is that, knowing $n_0$ from observations and $n_1$ from theoretical models for \methanol\ masers, our observed values of \Blos, and hence $B_1$, allow us to get a measure of the large-scale magnetic field $B_0$.

Our detected value of \Blos\ in the 44 GHz Class I \methanol\ maser line toward OMC-2 is similar to the fields ($\sim 25 $ mG) detected in our recent discovery of the Zeeman effect in the 36 GHz Class I \methanol\ line toward the high mass star forming region M8E (\citealt{sm2009}). It is also similar to the fields obtained from the 6.7 GHz Class II \methanol\ maser line observed by \citet{wv2008}, who detected significant magnetic fields with the 100 m Effelsberg telescope in this line toward 17 sources, with an average value of 23 mG. The possibility of a fake Zeeman pattern due to beam squint is brought up every time there is a new Zeeman detection, but it is unlikely in our maser observations; this is discussed in more detail in \citet{sm2009}. Since Class I and Class II \methanol\ masers likely trace different spatial regions (\citealt{elling2005}, and references therein), the similarities in the fields measured in these masers suggests that \methanol\ masers may trace the larger scale magnetic fields in star forming regions. Another possibility might be that Class I masers occur in the very early stages of star formation (before the formation of an ultracompact \ion{H}{2}\ region), and Class II masers occur later on. In that case, the similarity in \Blos\ for these two classes may indicate that the magnetic field strength remains the same during the early stages of the star formation process. It is possible, however, that the similarities may result simply from selection effects due to orientation and/or the shock process. Therefore, a larger statistical sample as well as measurements of the Land$\acute{e}$ $g$-factor are needed in order to test possible correlations or anti-correlations between fields measured in Class I and II masers and at different frequencies within each of these types. 

Finally, we use our observed value for the magnetic field to compare the magnetic and dynamical energies in these masing regions. The magnetic energy density is given by $3 \Blos^2/8\pi$. Here, we have used the relation $B^2 = 3 \Blos^2$ that was obtained by \citet{rmc99} on statistical grounds. For \Blos\ $\approx$ 18 mG from these observations, the magnetic energy density will be $4 \times 10^{-5}$ ergs \cm{-3}. The kinetic energy density (thermal and turbulent) is given by $(3/2)mn\sigma^2$, where
$\sigma = \Delta v/(8 \mathrm{ln} 2)^{1/2}$ is the velocity dispersion, the mass $m = 2.8\, m_\text{p}$ (assuming 10\% He; $m_\text{p}$ is the proton mass). Since masers occur only in special directions along which they have developed the required velocity coherence, the velocity dispersions in the masing region may be greater than that traced by masers. If we use $\Delta v = 5$ \kmS\ --- a reasonable value (see, e.g., \citealt{cfm99}), and $n = 10^6$ \cm{-3}, the density quoted in \citet{pp2008} at which maser action in the 44 GHz line is maximized, we get a kinetic energy density equal to $3 \times 10^{-7}$ ergs \cm{-3}. Even if $\Delta v$ is as high as 20 \kmS, as quoted by \citet{taka2008}\ for the CO (3-2) outflow marked at our \methanol\ maser position in Figure \ref{f850}, we will get  kinetic energy density equal to $5 \times 10^{-6}$ ergs \cm{-3}. Therefore, the magnetic energy is dominant by at least an order of magnitude, indicating that the magnetic fields are dynamically significant in these regions. Also of interest is that the magnetic energy appears to be dominant in different regions supporting Class I and Class II masers (e.g., this work; \citealt{sm2009}; \citealt{wv2008}).

\section{CONCLUSIONS}
\label{sCONC}

We have reported the discovery of the Zeeman effect in the 44 GHz Class I methanol (\methanol) maser line. We detected a magnetic field \Blos\  $= 18.4 \pm 1.1$ mG toward a star forming region in OMC-2. There may be systematic bias in this value due to the assumed Zeeman splitting factor of $z = 1.0$ Hz \mG, which is based on laboratory measurements in the 25 GHz \methanol\ maser line. Still, such values for \Blos\ imply that the magnetic energy may be dominant by at least an order of magnitude over the kinetic energy density, meaning that the magnetic field is dynamically significant in these regions. By assuming a feasible configuration for the shocks that give rise to 44 GHz \methanol\ masers, we find that our detected value of \Blos\ leads to a value for the density in the postshock region ($n_1 = 10^6$ \cm{-3}) that agrees with the value predicted by theoretical models for maximizing maser action in the methanol line  at 44 GHz. Alternatively, if we know the preshock density $n_0$ (e.g., from molecular line observations) and $n_1$ (from theoretical models of masers), we can use our detected \Blos\ to find the large-scale magnetic field. Together with the magnetic fields detected in the 36 GHz Class I \methanol\ maser line (\citealt{sm2009}), and 6.7 GHz Class II \methanol\ masers (\citealt{wv2008}), our observations reveal the potential for obtaining a more complete picture of the magnetic field in star forming regions, since Class I and Class II masers are found at different distances from the protostar, likely trace different physical conditions, and may be excited at different stages of the star formation process.  

Given that Class I and II methanol masers allow observation of star forming regions on small physical scales that are inaccessible to other lines, and the possibility that these masers may trace the large-scale magnetic field, we believe that our discovery motivates several observational approaches. A high angular resolution study of the association between 36 GHz and 44 GHz Class I methanol masers is now possible due to the new 36 GHz receivers on the EVLA. Likewise, a high angular resolution study of the association between Class I and Class II methanol masers is warranted. Identification of other maser sites where the Zeeman effect could be measured would allow for the conclusions above to be placed on a statistical footing. Moreover, the OMC-2 region that we have studied is forming intermediate stars (\citealt{taka2008}). Observations of magnetic fields in 44 GHz \methanol\ masers in high mass star forming regions, and the other studies described above, may yield important clues in the unsolved areas of high mass star formation. Finally, there is also a pressing need for laboratory measurements of the Zeeman splitting coefficients for various types of Class I and II \methanol\ masers. 

\acknowledgments
We would like to thank an anonymous referee whose suggestions have greatly improved the manuscript. We also thank Fr{\' e}d{\' e}rick Poidevin for sharing an eps file of Figure 7 from Poidevin et al. (2010). We have used extensively the NASA Astrophysics Data System (ADS) astronomy abstract service, and the astro-ph web server. 

\clearpage

%References

%\clearpage 

\begin{deluxetable}{lccccccrrrrrr}
\tablenum{1}
\tablewidth{0pt}
\tablecaption{Parameters for EVLA Observations \protect\label{tOP}}   
\tablehead{
\colhead{Parameter}  &  
\colhead{Value}}
\startdata
Observation Dates & 2009 Oct 25 \& Nov 25 \\
Configuration &   D \\
R.A.~of field center (J2000) & $05^\text{h} 35^\text{m} 27^\text{s}.66$ \\
Decl.~of field center (J2000) & $-$05$\arcdeg$09$\arcmin$39$\arcsec$.6  \\
Total Bandwidth & 1.56 MHz \\
No. of channels & 256 \\
Channel Spacing  & 0.04  \kmS \\
Total Observing Time  & 4 hr \\
Rest Frequency & 44.069488 GHz \\
Velocity at band center\tablenotemark{a} & 13.2 \kmS \\
Target source velocity & 11.6 \kmS \\
%Hanning Smoothing &  No \\
FWHM of synthesized beam & $1.93\arcsec \times 1.58\arcsec$  \\
& P.A. $= -10.40\arcdeg$ \\
Line rms noise\tablenotemark{b} &   8 mJy beam$^{-1}$
\enddata
%\tablenotetext{a}{We used a position intermediate between the Haystack position of this maser (\citealt{pp2008}), which has a 5 arcsec accuracy based on observations done between 2000 and 2005 (no specific date is given for their OMC-2 observations), and the position reported by \citet{sk2009} based on VLA observations carried out in 1998 with a positional accuracy of 0.01 arcsec.}
\tablenotetext{a}{The line was centered in the second half of the 1.56 MHz band in
order to avoid aliasing (see \S\ \ref{sODR}).}
\tablenotetext{b}{The line rms noise was measured from the stokes $I$ 
image cube using maser line free channels.}
\end{deluxetable}
     
\clearpage

\end{document}